\documentclass[aps,pre,twocolumn,showpacs]{revtex4}
\begin{document} \title{An evolution equation of the population genetics:
relation to the density-matrix theory of quasiparticles with
general dispersion laws}

\author{V. Bez{\' a}k}
\email{bezak@fmph.uniba.sk} \affiliation {Department of Solid
State Physics, Comenius University, 842 48 Bratislava, Slovakia}

\begin{abstract}
The Waxman-Peck theory of the population genetics is discussed in
regard of soil bacteria. Each bacterium is understood as a carrier
of a phenotypic parameter $p$. The central aim is the calculation
of the probability density with respect to $p$,
${\mathit\Phi}(p,t;p_0)$, of the carriers living at time $t>0$,
provided that initially, at $t_0=0$, all bacteria carried the
phenotypic parameter $p_0=0$. The theory involves two small
parameters: the mutation probability $\mu$ and a parameter
$\gamma$ involved in a function $w(p)$ defining the fitness of the
bacteria to survive the generation time $\tau$ and give birth to
offspring. The mutation from a state $p\:$ to a state $q$ is
defined by a Gaussian with a dispersion $\sigma_{\textrm{m}}^{2}$.
The author focuses attention on a function $\varphi(p,t)$ which
determines uniquely the function ${\mathit\Phi}(p,t;p_0)$ and
satisfies a linear equation (Waxman's equation). The Green
function of this equation is mathematically identical with the
one-particle Bloch density matrix where $\mu$ characterizes the
order of magnitude of the potential energy. (In the
$x$-representation, the potential energy is proportional to the
inverted Gaussian with the dispersion $\sigma_{\textrm{m}}^{2}$.)
The author solves Waxman's equation in the standard style of a
perturbation theory and discusses how the solution depends on the
choice of the fitness function $w(p)$. In a sense, the function
$c(p)=1-w(p)/w(0)$ is analogous to the dispersion function $E(p)$
of fictitious quasiparticles. In contrast to Waxman's
approximation where $c(p)$ was taken as a quadratic function,
$c(p)\:\approx\:\gamma p^2$, the author exemplifies the problem
with another function, $c(p)=\gamma[1-\exp(-\:ap^2)]$, where
$\gamma$ is small but $a$ may be large. The author shows that the
use of this function in the theory of the population genetics is
the same as the use of a non-parabolic dispersion law $E=E(p)$ in
the density-matrix theory. With a general function $c(p)$, the
distribution function ${\mathit\Phi}(p,t;0)$ is composed of a
delta-function component, $N(t)\delta(p)$, and a blurred
component. When discussing the limiting transition for
$t\:\to\:\infty$, the author shows that his function $c(p)$
implies that $N(t)\:\to\:N(\infty)\:\neq\:0$ in contrast with the
asymptotics $N(t)\:\to\:0$ resulting from the use of Waxman's
function $c(p)\:\sim\:p^2$.
\end{abstract}


\maketitle

\section{Introduction}

\bigskip
During the past century, the development of the quantum theory has
been paralleled with the development of the theory of stochastic
processes. When comparing the time-dependent Schr{\" o}dinger
equation of the quantum mechanics with the Fokker-Planck equation
of the stochastic dynamics, we may assert that both these
equations are of the same kind. Mathematically speaking, both
these equations are linear second-order partial differential
equations of the parabolic type. The Fokker-Planck equation can be
derived for any process which can be described by the Langevin
equation $\eta{\mbox{d}}x(t)/{\mbox{d}}t-F(x(t))=f(t)$ (cf. e.g.
[1, 2]). In the Langevin equation, $\eta>0$ is a deterministic
constant, $F(x)$ is a deterministically defined real-valued
function and $f(t)$ is a stochastically defined zero-centered
stationary Gaussian white-noise function. (As a rule, $x(t)$ and
$f(t)$ are considered as real random functions of the real time
variable $t$. The stationarity of $f(t)$ is meant in the
stochastic sense.) Using the angular brackets $\langle\ \rangle$
for the averaging with respect to the randomness of $f(t)$, we
assume that $\langle f(t)\rangle=0$ and $\langle
f(t_1)f(t_2)\rangle={\mathit\Lambda}\:\delta(t_1-t_2)$ at all time
instants $t$, $t_1$, $t_2$. The initial value of $x(t)$,
$x(0)=x_0$, is usually a deterministic value given in advance. The
Fokker-Planck equation concerns the conditional probability
density $P(x,t|x_0)=\langle\delta(x(t)-x)\rangle\:\geq\:0$:
\begin{equation}\label{FPeq}\frac{\partial P(x,t|x_0)}{\partial
t}=\frac{{\mathit\Lambda}}{2\eta^2}\:
\frac{\partial^2P(x,t|x_0)}{\partial x^2}\:-\:\frac{1}{\eta}\:
\frac{\partial}{\partial
x}{\bigg[}F(x)P(x,t|x_0){\bigg]}\:,\end{equation}
\begin{equation}\label{FPinit}P(x,+0|x_0)=\delta(x-x_0)\:.
\end{equation} If $x(t)$ is the
instantaneous position of a Brownian particle moving along a line,
we may speak of the diffusion coefficient $D$ (such that
$2\eta^2D={\mathit\Lambda}>0$) and of the mobility $1/\eta$ of the
particle. Then $F(x)=-\:{\mbox{d}}V_{\textrm{B}}(x)/{\mbox{d}}x$
is a driving force, $f(t)$ is the Langevin stochastic force and
(\ref{FPeq}) is the governing equation of the Brownian dynamics.
Equation (\ref{FPeq}) can be transformed into an equation of the
Schr{\" o}dinger type by the substitution
\begin{equation}\label{FPsubst}P(x,t|x_0)=
\exp{\bigg(}\frac{\eta}{{\mathit\Lambda}}\:
[V_{\textrm{B}}(x_0)-V_{\textrm{
B}}(x)]{\bigg)}\:R(x,t|x_0)\:.\end{equation} The function
$R(x,t|x_0)$ obeys the equation
\begin{equation}\label{SchrFP}\frac{\partial R(x,t|x_0)}{\partial
t}=\frac{{\mathit\Lambda}}{2\eta^2}\:
\frac{\partial^2R(x,t|x_0)}{\partial
x^2}\:-\:V(x)R(x,t|x_0)\:,\end{equation} where $$ V(x)=
\frac{1}{2\eta}\:{\bigg[}\frac{1}{{\mathit\Lambda}}\:
{\bigg(}\frac{{\mbox{d}}V_{\textrm{B}}(x)}{{\mbox{
d}}x}{\bigg)}^2\:-\: \frac{{\mbox{d}}^{2}V_{\textrm{B}}(x)}{{\mbox
{d}}x^{2}}{\bigg]}$$
\begin{equation}\label{VtoVB}=\:\frac{1}{2\eta}\:
{\bigg[}\frac{[F(x)]^2}{{\mathit\Lambda}}\:+\:
\frac{{\mbox{d}}F(x)}{{\mbox{d}}x}{\bigg]}\:.\end{equation}
Equation (\ref{SchrFP}) is known as the real Schr{\" o}dinger
equation. If $t$ is replaced by $\hbar\beta$, where $k_{\textrm
{B}}T=1/\beta$, equation (\ref{SchrFP}), with some change of
symbols, becomes the Bloch equation for the one-particle canonical
density matrix of boltzons of a constant (effective) mass in the
thermodynamic equilibrium at the temperature $T$ [3]. Thus, when
comparing equations (\ref{FPeq}) and (\ref{SchrFP}), we can always
juxtapose the Brownian dynamics and the quantum theory. (This
juxtaposition can also be based on Feynman's path-integral theory
[4, 5].) The transition from the formalism of the Brownian theory
to the formalism of the quantum theory is easy since if we define
the driving force
$F(x)=-{\:}{\mbox{d}}V_{\textrm{B}}(x)/{\mbox{d}}x$, we can
directly calculate the function $V(x)$ according to equation
(\ref{VtoVB}). ({\textit{Vice versa}}, if one tries to find $F(x)$
corresponding to a given function $V(x)$, one has to solve
equation (\ref{VtoVB}) which is nonlinear. Equation (\ref{VtoVB})
is known as the Riccati equation. Cf. any handbook on nonlinear
differential equations $-$ e.g. [6]. Recently the usefulness of
the Riccati equation in solving various problems of classical and
quantum mechanics has been widely corroborated [$7-11$].
Nonetheless, except for the rare possibility to derive analytical
solutions $F(x)$ of this equation in some cases when $V(x)$ is
chosen in a very simple and very special form, equation
(\ref{VtoVB}) cannot be solved otherwise than numerically. Thus,
the problem of finding a Brownian model to a given
quantum-mechanical model is relatively difficult.)

In evolutionary theories of various populations, we may use
$x(t)=\ln n(t)-\overline{\ln n(t)}$, taking $n(t)$ as the number
of individuals of a certain kind at the time instant $t$ and
defining $\overline{\ln n(t)}$ as an average value of $\ln n(t)$.
Since $n(t)$ may represent very large numbers, $n(t)$ may be
treated as a continuous function so that the values of $x(t)$ may
span the whole set of real numbers. Since mutations are random
events, a formal stochastic theory of the population genetics can
certainly be based on the use of the Langevin equation and,
correspondingly, of the Fokker-Planck equation (\ref{FPeq}). If it
is advantageous, we may also use the equivalent
Schr{\"o}dinger-type equation (\ref{SchrFP}).

Notwithstanding, recently Waxman has shown that there is also
another mathematical relation between the population genetics and
the quantum mechanics [12]. Waxman's theory concerns a simplified,
but well-founded, model that we call the Waxman-Peck model. (Cf.
[13, 14] and references quoted therein.) Waxman's
Schr{\"o}dinger-type equation involves, in the $x$-representation,
the `potential energy' function
\begin{equation}\label{Waxpot}V(x)\:\sim\:-\:
\mu\:\exp{\bigg(}-\:\frac{\sigma_{\textrm {m}}^2x^2}{2}{\bigg)}
\end{equation}
with two mutation parameters, $\mu>0$ and $\sigma_{\textrm{m}}>0$.
The `kinetic-energy' operator ${\hat T}$ in Waxman's equation was
chosen in the usual form, ${\hat
T}\:\sim\:-\:\partial^2/{\partial}x^2$.

In the present paper, we will generalize Waxman's theory. In the
momentum representation, Waxman's `kinetic energy' is quadratic,
$\langle p|{\hat T}|p\rangle\:\sim\:p^2$, as it is in the quantum
mechanics. On the other hand, we assume that $\langle p|{\hat
T}|p\rangle$, being a positive function, need not be quadratic; we
only require its analyticity along the real $p$-axis. This
variability offers further possibilities to model the genetic
evolution by adequate fitness functions (cf. Section II).
Obviously, if ${\langle}p|{\hat T}|p\rangle>0$ is non-quadratic,
the mathematical relation between the evolution equation of the
population genetics and the quantum theory becomes somewhat more
sophisticated than in Waxman's case. Namely, when paying heed to
the $x$-representation, we have generally to consider a more
complicated equation than the real Schr{\"o}dinger equation: in
general, our equation, with the Hamilton operator $E({\hat
p})+V(x)$, where $E(p)\:\sim\:c(p)$, is a {\textit{functional}}
differential equation. (Note that ${\hat p}=-\:{\mbox
i}\partial/\partial x$. The McLaurin development of $c(p)$ may
involve an infinite number of terms.) For solid state theorists,
such an equation is familiar as the transformed effective mass
equation (cf. e.g. the monograph [15] or our paper [16]). This
equation was invented for envelope wave functions of electrons in
crystalline solids. (Synonymously, we may also speak of the real
Schr{\"o}dinger-Wannier equation. It is identical with the
`one-particle Bloch equation' for the canonical density-matrix
with the Hamiltonian $E({\hat p})+V(x)$.)

From the viewpoint of the effectiveness of calculations in the
present paper, we deem the momentum representation better than the
$x$-representation. Under the assumption of the smallness of the
parameter $\mu$, we can apply the `plane-wave perturbation theory'
(Section III) of the density-matrix theory. For a broad class of
fitness functions, the distribution function of the theory of the
population genetics can be expressed as a linear expression of the
mutation parameter $\mu$.

\section{The Waxmann-Peck model of the population genetics}

\bigskip
Let us assume that a large enough habitat (such as a given volume
of soil) hosts bacteria of a certain kind. The habitat yields
space, food, moisture, temperature, inhibitory substances and
other needs for the survival of the bacteria in the sense that the
total number of the bacteria will never decrease to zero and will
never increase to infinity. Most of bacteria are free-living
microorganisms multiplying by simple fission. This means that
their reproduction is asexual. In other words, each bacterial
individual has only one parent. The typical number of bacteria may
be huge indeed: one gram of soil may contain several hundreds
million bacteria. Although the bacteria are small organisms $-$
usually 0.3 - 2 micrometres in diameter $-$ their morphology is
well distinguishable microscopically.

There are two most frequent shapes of soil bacteria: short rods
and (slightly deformed) spheres. In both these cases, we may
characterize each bacterium by its size $s$. For instance, if the
bacterium resembles a rod, we define $s$ as the length of the rod.
Denoting the average of $\ln s$ as $\overline{\ln s}$, we define
the phenotypic parameter as $p=\ln s-\overline{\ln s}$. Recalling
to biology, we consider the phenotypic parameter as an inheritable
value. If there were no mutations in the reproduction of the
bacteria, all the bacterial individuals in each generation would
be equally long, i.e. $p$ would be a constant equal to $p_0=0$.
But then, the mutations $-$ however infrequent they may be $-$
account for a dispersal of the value $p$ among the individuals,
despite the fact that all individuals under consideration do still
belong to the same biological type.

Generally, the theory has to respect both mutations caused by
environmental effects and spontaneous mutations. For the sake of
simplicity, we will consider no other than the spontaneous
mutations. The spontaneous mutations are mainly due to
transcription mistakes in the replication of the DNA, i.e. in the
transmission of the genetic information just at the reproduction
events. Let $\mu>0$ be the probability of the occurrence of such a
mutation. In the case of soil bacteria, biologists have estimated
the values of $\mu$ between $10^{-8}$ and $10^{-5}$. If a mother
bacterium is the carrier of the phenotypic value $p$, the daughter
bacterium will carry the same value $p$ with the probability equal
to $1-\mu$. (We assume that a new-born daughter bacterium grows
quickly enough to the adult size before becoming mature so that we
need not distinguish between the size of young and adult
bacteria.) If the birth of the daughter bacterium is accompanied
with a mutation of the DNA, then there is a non-zero probability
$M(q-p){\mbox {d}}q$ for the possibility that the phenotypic value
$q$ of the daughter bacterium may lie in the interval
$(q,q+{\mbox{d}}q)$, provided that the phenotypic value of the
mother bacterium was equal to $p$. Following Waxman, we take the
function $M(p)$ as a Gaussian,
\begin{equation}\label{Gaussmut}M(p)\:\equiv\:M(p;\sigma_{\textrm{m}})=
{\bigg(}\frac{1}{2\pi\sigma_{\textrm{m}}^2}{\bigg)}^{1/2}
\exp{\bigg(}-\:\frac{p^2}{2\sigma_{\textrm
{m}}^2}{\bigg)}\:.\end{equation} Here $\sigma_{\textrm{m}}^2$ is
the dispersion of values of $p$.

Now, to formulate the evolution equation of the population
genetics, we have to introduce the average generation time $\tau$.
Simplifying the problem, we may consider a discrete time variable
as follows. Let the births of the bacteria happen at the time
instants $t_n=(n-1)\tau$, $n=1,2,\dots$ Then we may say that the
bacteria of the $n$th generation live between $t_{n-1}$ and $t_n$.
Thus, $n$ is the generation index. The time discretization is an
auxiliary, rather formal, mathematical trick which loses its
significance if the time $t$ is continualized. For each $n$, we
define the distribution function ${\mathit\Phi}_n(p)$ so that
${\mathit\Phi}_n(p){\mbox{d}}p$ may be interpreted as the
probability of the occurrence of the phenotypic value $p$ in the
interval $(p,p+{\mbox{d}}p)$ in the $n$th generation. The basic
problem is to relate the distribution function of the generation
number $n+1$ (`generation of daughters') with the distribution
function of the generation number $n$ (`generation of mothers').

Before writing the recurrent formula between the functions
${\mathit\Phi}_{n+1}(p)$ and ${\mathit\Phi}_n(p)$, which is our
primary goal in this section, we have still to mention one
important point. Even if we have neglected the environmental
influence upon the mutations, we do have to consider environmental
effects in a Darwinian sense. Namely, we have to respect that not
all bacteria, after their birth, are equally fit to survive over
the whole generation time $\tau$. Only those bacteria whose age is
equal to $\tau$ give birth to offspring. Some of the bacteria die
before becoming mature. These bacteria do not take part in
producing the individuals of the next generation. (However, we
assume that even the fittest mother bacterium dies soon after
giving birth to the daughter bacterium. Therefore, we do not
include the mother bacteria in the number of the bacteria living
in the time interval $(t_n,t_{n+1})$. The mother bacteria have
been included in the number of the bacteria living in the time
interval $(t_{n-1},t_n)$.) The fitness of the bacteria to live in
their environment until their maturity can be modeled by a
non-negative function $w(p)$. Requiring that
\begin{equation}\label{fitness}0<w(p)<1\:,\end{equation}
we may give the function $w(p)$ a probabilistic meaning. We assume
that a new-born carrier of the phenotypic value $p$ has the chance
to live until the maturity with the probability $w(p)$. The number
of mature carriers of the phenotypic value $p$ from the interval
$(p,p+{\mbox{d}}p)$ in the $n$th generation is proportional to
$w(p){\mathit\Phi}_n(p)$. To determine the shape of the function
$w(p)$ should be a matter of thorough biological investigations
from case to case. We suppose, as Waxman and Peck did, that $w(p)$
behaves analytically around the value $p_0=0$ and that this value
corresponds to the maximum value of $w(p)$. (Apparently, the
fittest bacterial individuals are those whose phenotypic parameter
$p$ is equal to the average value $\bar p$. However, $\bar p=0$.)

Waxman and Peck have chosen the function $w(p)$ in the special
form $$w(p)=w(0)\exp(-\:\gamma p^2)\:,\ \
0<w(0)<1\:,\eqno(8\mbox{a})$$ assuming that $0<\gamma\:\ll\:1$.
(In fact, expression (8a) {\textit{defines}} the Waxman-Peck
model. The value of $w(0)$ is insignificant since the distribution
functions ${\mathit\Phi}_n(p)$ are independent of $w(0)$.)

There are, of course, many other possibilities to model $w(p)$ by
slowly varying functions with the maximum at $p_0=0$. These
functions need not tend to zero if $|p|\:\to\:\infty\:$. (The
fitness function has been defined as a {\textit{probability}}, not
as a probability density!) In order to illustrate how the theory
may depend on the choice of the function $w(p)$, we will treat, in
addition to the Waxman-Peck model, also an alternative model. Our
model (considered as an example) is defined by the fitness
function $$w(p)=w(0)\:{\lbrace}1\:-\:
\gamma[1-\exp(-\:ap^2)]{\rbrace}\:.\eqno(8\mbox{b})$$ Here we
suppose that $0<\gamma\:\ll\:1$, admitting that $a>0 $ need not be
a small number. Expression (8b) tends to $w(0)(1-\gamma)>0$ if
$|p|\:\to\:\infty$.

The distribution function ${\mathit\Phi}_{n+1}(p)$ of the
generation of daughters is determined by two contributions from
the generation of mothers. The first stems from the births without
mutations. The phenotypic value $p$ is unchanged at such births
and the probability of the occurrence of such births is equal to
$1-\mu$. The first contribution to ${\mathit\Phi}_{n+1}(p)$ is
proportional to $(1-\mu)w(p){\mathit\Phi}_n(p)$. The second
contribution to ${\mathit\Phi}_{n+1}(p)$ is proportional to
$\mu\int_{-\infty}^\infty{\mbox{d}}q\:M(p-q;\sigma_{\textrm
{m}})w(q){\mathit\Phi}_n(q)$. The interpretation of this
expression is clear: if the birth of the carrier of the phenotypic
value $p$ is accompanied with a mutation, we have to consider
mature individuals, allowing all possible phenotypic values $q$ of
potential mothers. To exhaust all such possibilities, we have to
integrate $M(p-q;\sigma_{\textrm{m}})w(q){\mathit\Phi}_n(q)$ with
respect to $q$. Since both ${\mathit\Phi}_{n}(p)$ and
${\mathit\Phi}_{n+1}(p)$ are probability densities, we have to
require that
\begin{equation}\label{normalization}\int_{-\infty}^\infty{\mbox
{d}}p\:{\mathit\Phi}_n(p)= \int_{-\infty}^\infty{\mbox
{d}}p\:{\mathit\Phi}_{n+1}(p)=1\:.\end{equation} Therefore, we
write the equality

\bigskip
${\mathit\Phi}_{n+1}(p)=$
\begin{equation}\label{recur}
\frac{(1-\mu)w(p){\mathit\Phi}_n(p)+\mu\int_{-\infty}^\infty{\mbox
{d}}q\: M(p-q;\sigma_{\textrm{m}})
w(q){\mathit\Phi}_n(q)}{\int_{-\infty}^\infty{\mbox
{d}}q\:w(q){\mathit\Phi}_n(q)}\:.\end{equation} The denominator in
the r.h. side of equation (\ref{recur}) warrants the fulfillment
of condition (\ref{normalization}). Since $\gamma$ is small, it is
convenient to introduce the complementary function $c(p)$ to
$w(p)/w(0)$:
\begin{equation}\label{cWax}c(p)=1-\frac{w(p)}{w(0)}\:.\end{equation}
In the case of the Waxman-Peck model, $$c(p)=1-\exp(-\:\gamma
p^2)\:,\eqno(11{\mbox {a}})$$ whilst in the case of the model
defined by function (8b),
$$c(p)=\gamma\:[1-\exp(-\:ap^2)]\:.\eqno(11{\mbox{b}})$$ From the
viewpoint of biology, the smallness of $\gamma$ implies that the
comparison of the survival fitness of the majority of the bacteria
with the survival fitness of the fittest bacteria should not
reveal too conspicuous differences. When using the function
$c(p)$, we can rewrite formula (\ref{recur}) in the form
\begin{widetext}
$${\mathit\Phi}_{n+1}(p)=\frac{(1-\mu)[1-c(p)]{\mathit\Phi}_n(p)+\mu
\int_{-\infty}^\infty{\mbox{d}}q\:M(p-q;\sigma_{\textrm
{m}})[1-c(q)]{\mathit\Phi}_n(q)} {1-\int_{-\infty}^\infty{\mbox
{d}}q\:c(q){\mathit\Phi}_n(q)}\:.\eqno(10')$$
\end{widetext}
With realistic values of $p$ around $p_0=0$, the values of $\gamma
p^2$ are small. Thus, in the case of the Waxman-Peck model, the
values of $c(p)$ are also small and $$c(p)=\gamma p^2+{\mathcal
{O}}(\gamma^2)\:.\eqno(11'{\mbox{a}})$$ On the other hand, in the
case of the model defined by function (8b), we have to keep
expression (11b) intact since $a$ need not be a small parameter.

We may take advantage of the possibility to neglect all terms of
the order of magnitude of $\gamma^2$, as well as of $\gamma\mu$.
So we write $$\frac{1}{1-\int_{-\infty}^\infty{\mbox
{d}}q\:c(q){\mathit\Phi}_n(q)}= 1+\int_{-\infty}^\infty{\mbox
{d}}q\:c(q){\mathit\Phi}_n(q)+\dots$$ and
\begin{widetext}
$${\mathit\Phi}_{n+1}(p)={\mathit\Phi}_n(p)\:-\:{\bigg[}c(p)\:-\:
\int_{-\infty}^\infty{\mbox
{d}}q\:c(q){\mathit\Phi}_n(q){\bigg]}{\mathit\Phi}_n(p)\:
-\:\mu\:{\bigg[} {\mathit\Phi}_n(p)-\int_{-\infty}^\infty{\mbox
{d}}q\: M(p-q;\sigma_{\textrm
{m}}){\mathit\Phi}_n(q){\bigg]}+\dots\eqno(10'')$$
\end{widetext}
Now, in the approximation neglecting the terms symbolized by the
dots, we are ready to go over into the formalism employing the
continual time variable $t$, realizing that the value of the
generation index $n$ may be high. Typically, the generation time
$\tau$ of soil bacteria is about 20 minutes. This means that after
elapsing hundred days, the genetic information passes over more
than seven thousand generations of the bacteria. If $n\:\gg\:1$,
we may identify ${\mathit\Phi}_n(p)$ with ${\mathit\Phi}(p,t)$ and
approximate the difference
${\mathit\Phi}_{n+1}(p)-{\mathit\Phi}_n(p)$ as the time
derivative:
\begin{equation}\label{timederiv}{\mathit\Phi}_{n+1}(p)-{\mathit\Phi}_n(p)=\tau\:
\frac{\partial{\mathit\Phi}(p,t)}{\partial
t}\:+\:\dots\end{equation} Thus, we can rewrite equation ($10''$)
in the approximate integro-differential form
\begin{widetext}
\begin{equation}\label{Waxnonlin}
\frac{\partial{\mathit\Phi}(p,t)}{\partial t}=
-\:\frac{1}{\tau}\:{\bigg[}c(p)- \int_{-\infty}^\infty{\mbox
{d}}q\:c(q){\mathit\Phi}(q,t){\bigg]}{\mathit\Phi}(p,t)\:
-\:\frac{\mu}{\tau}\:{\bigg[}
{\mathit\Phi}(p,t)-\int_{-\infty}^\infty{\mbox{d}}q\:
M(p-q;\sigma_{\textrm{m}}){\mathit\Phi}(q,t){\bigg]}\:.\end{equation}
\end{widetext}
This equation was derived in [12] (where, however,$c(p)$ was
approximated as $\gamma p^2$). Evidently, equation
(\ref{Waxnonlin}) is nonlinear. Fortunately, this nonlinearity
does not mean a serious problem since we may employ the
substitution
\begin{equation}\label{Waxsubst}{\mathit\Phi}(p,t)=\frac{\varphi(p,t)}
{\int_{-\infty}^\infty{\mbox{d}}q\:\varphi(q,t)}\:,\end{equation}
and require the validity of the equation
\begin{equation}\label{Waxdiflin}
\frac{\partial\varphi(p,t)}{\partial t}=-\:
\frac{1}{\tau}\:c(p)\varphi(p,t)\:+\:
\frac{\mu}{\tau}\int_{-\infty}^\infty{\mbox{d}}q\:
M(p-q;\sigma_{\textrm{m}})\varphi(q,t)\:.\end{equation} Equation
(\ref{Waxdiflin}) is linear. After integrating it with respect to
$p$, we obtain the equation
$\mbox{d}/{\mbox{d}}t\int_{-\infty}^\infty{\mbox{d}}q\:\varphi(q,t)=
-\:(1/\tau)\int_{-\infty}^\infty{\mbox
{d}}p\:c(p)\varphi(p,t)\:+\:
(\mu/\tau)\int_{-\infty}^\infty{\mbox{d}}q\:\varphi(q,t)$ and when
substituting expression (\ref{Waxsubst}) for $\varphi(p,t)$, we
arrive at the identity $$\frac{\mbox{d}}{{\mbox
{d}}t}\int_{-\infty}^\infty{\mbox{d}}q\:\varphi(q,t)=$$
\begin{equation}\label{derivtotal}
\int_{-\infty}^\infty{\mbox{d}}q\:\varphi(q,t)\:{\bigg[}
-\:\frac{1}{\tau}\int_{-\infty}^\infty{\mbox
{d}}p\:c(p){\mathit\Phi}(p,t)\:+\:
\frac{\mu}{\tau}{\bigg]}\:.\end{equation} The differentiation of
expression (\ref{Waxsubst}) gives the identity
$$\frac{\partial\varphi(p,t)}{\partial t}=$$
\begin{equation}\label{derivmix}\frac{\partial{\mathit\Phi}(p,t)}{\partial
t}\:\int_{-\infty}^\infty{\mbox {d}}q\:\varphi(q,t)\:
+\:{\mathit\Phi}(p,t)
\:\frac{\mbox{d}}{{\mbox{d}}t}\int_{-\infty}^\infty{\mbox
{d}}q\:\varphi(q,t)\:.\end{equation} When equalizing the r.h.
sides of equations (\ref{Waxdiflin}) and (\ref{derivmix}) and when
respecting identity (\ref{derivtotal}), we obtain equation (l3)
for the function ${\mathit\Phi}(p,t)$. Thus, instead of directly
solving equation (\ref{Waxnonlin}), we may solve Waxman's equation
(\ref{Waxdiflin}) at first. This task, as we will show in Section
III, is not difficult. If the function ${\mathit\Phi}(p,t)$ obeys
linear boundary conditions, the function $\varphi(p,t)$ has to
obey the same boundary conditions. We will simply assume that
\begin{equation}\label{asymptot}{\mathit\Phi}(p,t)\:\to\:0\ \textrm{ and }\
\varphi(p,t)\:\to\:0\ \textrm{ if }\
|p|\:\to\:\infty\:.\end{equation} It remains still to discuss the
initial condition. Whichever initial function
\begin{equation}\label{initfunct}
{\mathit\Phi}(p,0)={\mathit\Phi}_0(p)\end{equation} is chosen, the
solution ${\mathit\Phi}(p,t)$ for $t>0$ of equation
(\ref{Waxnonlin}) is unique. Since equation (\ref{Waxdiflin}) is
linear, we may multiply $\varphi(p,t)$ by an arbitrary constant
$A$. If $\varphi(p,t)$ gives the function ${\mathit\Phi}(p,t)$
then $A\varphi(p,t)$ does also give the same function
${\mathit\Phi}(p,t)$. Therefore, we may choose the integral
$\int_{-\infty}^\infty{\mbox {d}}q\:\varphi(q,0)$ (which is a
constant) equal to unity. Then formula (\ref{Waxsubst}) and
equality (\ref{initfunct}) give us the initial condition
\begin{equation}\label{initcond}{\varphi}(p,0)=
{\mathit\Phi}_0(p)\end{equation} for the function $\varphi(p,t)$.

If ${\mathit\Phi}_0(p)$ is an even function, equation
(\ref{Waxdiflin}) implies that the function $\varphi(p,t)$ is also
even in the variable $p$ and
${\mathit\Phi}(-p,t)={\mathit\Phi}(p,t)$ at all times $t>0$. In
this case, the mean value of $p$ is an invariant in time (i.e. a
constant): \begin{equation}\label{barp}{\bar
p}=\int_{-\infty}^\infty{\mbox {d}}p\
p\:{\mathit\Phi}(p,t)=0\:.\end{equation}

Equation (\ref{Waxdiflin}) is formally the same as the
Schr{\"o}dinger-Wannier equation in the momentum representation.
It can easily be Fourier-transformed. We define the function
\begin{equation}\label{Fourierdir}\psi(x,t)=
\frac{1}{\sqrt{2\pi}}\:\int_{-\infty}^\infty{\mbox {d}}p\:
\exp({\mbox{i}}px)\varphi(p,t)\:.\end{equation} This function is
the solution of the functional differential equation
$$\frac{\partial\psi(x,t)}{\partial t}=-\:\frac{1}{\tau}\:
c{\bigg(}-\:{\mbox{i}}\:\frac{\partial}{\partial
x}{\bigg)}\psi(x,t)$$
\begin{equation}\label{functeq}
+\:\frac{\mu}{\tau}\: \exp{\bigg(}-\:\frac{\sigma_{\textrm
{m}}^2x^2}{2}{\bigg)}\psi(x,t)\:.\end{equation} In Waxman's
approximation, equation (\ref{functeq}) reads:
$$\frac{\partial\psi(x,t)}{\partial t}=\frac{\gamma}{\tau}\:
\frac{\partial^2\psi(x,t)}{\partial x^2}\:$$
$$+\:\frac{\mu}{\tau}\: \exp{\bigg(}-\:\frac{\sigma_{\textrm
{m}}^2x^2}{2}{\bigg)}\psi(x,t)\:. \eqno(23{\mbox{a}})$$ If $c(p)$
is taken in the form of expression (11b), the functional
differential equation for $\psi(x,t)$ reads
$$\frac{\partial\psi(x,t)}{\partial t}= \frac{\gamma}{\tau}\:
{\bigg[}\exp{\bigg(}a\frac{\partial^2}{\partial
x^2}{\bigg)-1}{\bigg]} \psi(x,t)\:$$
$$+\:\frac{\mu}{\tau}\:\exp{\bigg(}-\:\frac{\sigma_{\textrm
{m}}^2x^2}{2}{\bigg)}\psi(x,t)\:. \eqno(23{\mbox{b}})$$

\section{The plane-wave perturbation theory}

\bigskip
Instead of solving equation (23a) or equation (23b) and carrying
out the integration
\begin{equation}\label{Fourierinv}\varphi(p,t)=
\frac{1}{\sqrt{2\pi}}\:\int_{-\infty}^\infty{\mbox {d}}x\:
\exp(-\:{\mbox{i}}px)\psi(x,t)\:,\end{equation} we prefer to
calculate the function $\varphi(p,t)$ directly. Defining the
`potential-energy operator' ${\hat V}(p)$
\begin{equation}\label{hatV}{\hat V}(p)\varphi(p,t)=\frac{1}{\tau}\:
\int_{-\infty}^\infty{\mbox
{d}}q\:M(p-q;\sigma_{\textrm{m}})\varphi(q,t)\:, \end{equation}
let us write equation (\ref{Waxdiflin}) in the form
\begin{equation}\label{Waxeqhat}\frac{\partial\varphi(p,t)}{\partial
t}=-\:\frac{c(p)}{\tau}\:\varphi(p,t) \:+\:\mu\:{\hat
V}(p)\varphi(p,t)\end{equation} and define the Green function
$G(p,t;p_0)$ of this equation. Employing the Green function, we
write $\varphi(p,t)$ (for $t>0$) as the integral
\begin{equation}\label{intequ}\varphi(p,t)=\int_{-\infty}^\infty{\mbox
{d}}p_0\:G(p,t;p_0)\varphi(p_0,0)\:. \end{equation} The initial
function $\varphi(p,0)$ has been defined by equality
(\ref{initcond}). The Green function itself obeys the equation
\begin{equation}\label{integrodif}\frac{\partial G(p,t;p_0)}{\partial
t}=-\:\frac{c(p)}{\tau}\:G(p,t;p_0)\:+ \:\mu\:{\hat
V}(p)G(p,t;p_0)\:.\end{equation} According to equality
(\ref{intequ}), $G(p,t;p_0)$ satifies the initial condition
\begin{equation}\label{Greeninit}G(p,0;p_0)=\delta(p-p_0)\:.
\end{equation} In the special case when $c(p)$ is approximated by the
quadratic function, equation (\ref{integrodif}) is formally
identical with the Bloch equation for the one-particle canonical
density matrix $C_\beta(p,p_0)$ in the thermodynamic equilibrium.
In the case of a general function $c(p)$ we have to speak of
quasiparticles with a non-parabolic dispersion law. The function
$C_{+0}(p,p_0)$ is equal to $\delta(p-p_0)$ for quantum-mechanical
reasons. When transforming equation (\ref{integrodif}) into the
$x$-representation form, one observes that the `potential energy'
corresponds to a {\textit {well}}: it is an inverted Gaussian (cf.
expression (\ref{Waxpot})).

We can derive $G(p,t;p_0)$ as the series
\begin{equation}\label{Dysonseries}
G(p,t;p_0)=\sum_{j=0}^\infty\mu^j\:K_j(p,t;p_0)\:.\end{equation}
The zero-order term is the solution of the equation
\begin{equation}\label{G0eq}\frac{\partial G_0(p,t;p_0)}{\partial
t}=-\:\frac{c(p)}{\tau}\:G_0(p,t;p_0) \end{equation} with respect
to the condition
\begin{equation}\label{G0init}G_0(p,0;p_0)=\delta(p-p_0)\:.
\end{equation} When solving equation (\ref{G0eq}), we obtain, for
$t>0$, the function
\begin{equation}\label{G0solution}
G_0(p,t;p_0)=\delta(p-p_0)\exp{\bigg(}-\:\frac{c(p)t}{\tau}{\bigg)}\:.
\end{equation} With this function, we can write down the integral form of
equation (\ref{integrodif}):

\bigskip
$G(p,t;p_0)=G_0(p,t;p_0)\:$
\begin{equation}\label{inteqG}+\:\mu\:\int_0^t{\mbox
{d}}t_1\int_{-\infty}^\infty{\mbox{d}}p_1\:G(p,t-t_1;p_1) {\hat
V}(p_1)G_0(p_1,t_1;p_0)\:.\end{equation} This is a Dyson-type
series (cf. e.g. [17].) The first-order term in this series
(linear in $\mu$) reads:

\bigskip
$\mu G_1(p,t;p_0)=$
$$\mu\int_0^t{\mbox{d}}t_1\int_{-\infty}^\infty{\mbox
{d}}p_1\:G_0(p,t-t_1;p_1) {\hat V}(p_1)G_0(p_1,t_1;p_0)$$
$$=\frac{\mu}{\tau}
\int_0^t{\mbox{d}}t_1\int_{-\infty}^\infty{\mbox
{d}}p_1\:G_0(p,t-t_1;p_1)$$
\begin{equation}\label{Dyson1}\times\ \int_{-\infty}^\infty{\mbox
{d}}q\:M(p_1-q;\sigma_{\textrm{m}})G_0(q,t_1;p_0)\:.\end{equation}
After inserting expressions (\ref{Gaussmut}) and
(\ref{G0solution}) here, we obtain the function $$\mu
G_1(p,t;p_0)=\frac{\mu}{\tau}{\bigg(}\frac{1}{2\pi\sigma_{\textrm
{m}}^2}{\bigg)}^{1/2}
\exp{\bigg(}-\:\frac{(p-p_0)^2}{2\sigma_{\textrm{m}}^2}{\bigg)}\:$$
$$\times\ \int_0^t{\mbox{d}}t_1\:\exp{\bigg(}-\:
\frac{c(p)(t-t_1)+c(p_0)t_1}{\tau}{\bigg)} \:.$$ After performing
the integration with respect $t_1$, we arrive, respecting formula
(\ref{Gaussmut}), at the final result

\bigskip
$\mu G_1(p,t;p_0)=$
\begin{equation}\label{Dyson1final}\mu M(p-p_0;\sigma_{\textrm{m}})\:
\frac{\exp[-\:c(p_0)t/\tau]-\exp[-\:c(p)t/\tau]}{c(p)-c(p_0)}\:.
\end{equation}
In this same way, we could also calculate higher-order terms (i.e.
the terms proportional to $\mu^j$ with $j>1$) in series
(\ref{Dysonseries}). We expect, however, that higher-order terms
are negligible, since the mutation probability $\mu$ is, as
biologists have proved in their extensive studies, very small.

\bigskip
\section{Development of the phenotypic diversity in a population whose
individuals are initially equal}

\bigskip
If all individuals of a population are initially, at the time
$t_0=0$, carriers of the same phenotypic value $p_0$, the initial
distribution function ${\mathit\Phi}_0(p)$ is equal to the delta
function:
\begin{equation}\label{initequal}{\mathit\Phi}_0(p)=\delta(p)\:.
\end{equation} ($p_0={\bar p}=0$ according to our definition of the
phenotypic parameter $p$.) In regard to identity
(\ref{initcond}), equation (\ref{intequ}) allows us to assert that
\begin{equation}\label{initequal1}\varphi(p,t)=G(p,t;0)\:.
\end{equation} Since $c(0)=0$ (cf. expression (\ref{cWax})), formulae
(\ref{G0solution}) and (\ref{Dyson1final}) imply, respectively,
that
\begin{equation}\label{phaseSgen}G_0(p,t;0)=\delta(p)\end{equation} and
\begin{equation}\label{phaseGgen}\mu G_1(p,t;0)=
\mu M(p;\sigma_{\textrm{m}})\:
\frac{1-\exp[-\:c(p)t/\tau]}{c(p)}\end{equation} at all times
$t>0$. Hence, in the linear approximation with respect to $\mu$,
we have got the function
\begin{equation}\label{phigen}\varphi(p,t)= \delta(p)+\mu
M(p;\sigma_{\textrm {m}})\:\frac{1-\exp[-\:c(p)t/\tau]}{c(p)}\:.
\end{equation} The only problem that we have still left unsolved is the
calculation of the integral
$$\int_{-\infty}^\infty{\mbox{d}}p\:\varphi(p,t)=1\:+\:\mu\:
{\bigg(}\frac{1}{2\pi\sigma_{\textrm{m}}^2} {\bigg)}^{1/2}\
\times$$
\begin{equation}\label{intphi}\:\int_{-\infty}^\infty{\mbox{d}}p\:
\exp{\bigg(}-\:\frac{p^2}{2\sigma_{\textrm{m}}^2}{\bigg)}\:
\frac{1-\exp[-\:c(p)t/\tau]}{c(p)}\:.\end{equation} Recall that,
according to (\ref{Waxsubst}),
\begin{equation}\label{Phitophinorm}{\mathit\Phi}(p,t)=N(t)\varphi(p,t)\:,
\end{equation}
where
\begin{equation}\label{invnorm}N(t)={\bigg[}\int_{-\infty}^{\infty}
{\mbox{d}}p \:\varphi(p,t){\bigg]}^{-1}\end{equation} We will
calculate the function $N(t)$ approximately, assuming that
$0<\sigma_{\textrm {m}}<1$. (In fact, it is probable that
$\sigma_{\textrm {m}}\:\ll\:1$.) The most relevant values of $p$
contributing to the value of the integral in the r.h. side of
formula (\ref{intphi}) lie in the interval
$(-\:\sigma_{\textrm{m}},\sigma_{\textrm{m}})$.

\subsection{Distribution function ${\mathit\Phi}(p,t;0)$ in the
model where $c(p)=1-\exp(-\:\gamma p^2)$ $ ($the Waxman-Peck
model$)$}

\bigskip
Since $0<\gamma\sigma_{\textrm{m}}^2\:\ll\:1$, we may use, when
calculating integral (\ref{intphi}), the approximation expressed
by formula ($11'{\mbox{a}}$). Thus,
\begin{widetext}
$$\int_{-\infty}^\infty{\mbox
{d}}p\:\varphi(p,t)\:\approx\:1\:+\:\mu\:
{\bigg(}\frac{1}{2\pi\sigma_{\textrm{m}}^2}
{\bigg)}^{1/2}\:\int_{-\infty}^\infty{\mbox{d}}p\:
\exp{\bigg(}-\:\frac{p^2}{2\sigma_{\textrm{m}}^2}{\bigg)}\:
\frac{1-\exp(-\:\gamma p^2t/\tau)}{\gamma p^2}$$
$$=1\:+\:\frac{\mu}{\tau}\: {\bigg(}\frac{1}{2\pi\sigma_{\textrm
{m}}^2}{\bigg)}^{1/2}\:\int_0^t{\mbox{d}}t_1\:
\int_{-\infty}^\infty{\mbox
{d}}p\:{\bigg[}-\:\exp{\bigg(}\frac{1}{2\sigma_{\textrm
{m}}^2}\:+\:\frac{\gamma t_1}{\tau}{\bigg)p^2}{\bigg]}\:.$$
\end{widetext}
After carrying out the integration with respect to $p$, we obtain
the simple result $$\int_{-\infty}^\infty{\mbox
{d}}p\:\varphi(p,t)\:\approx\:1\:+\: \frac{\mu}{\tau}\:\int_0^t\:
\frac{{\mbox{d}}t_1}{(1+2\gamma\sigma_{\textrm
{m}}^2t_1/\tau)^{1/2}}$$
$$=\:1\:+\:\frac{\mu}{\gamma\sigma_{\textrm
{m}}^2}\:{\bigg[}{\bigg(}1\:+\: \frac{2\gamma\sigma_{\textrm
{m}}^2t}{\tau}{\bigg)}^{1/2}\:-\:1{\bigg]}\:.$$ Hence, according
to formula (\ref{Waxsubst}), we obtain the distribution function

\bigskip
${\mathit\Phi}(p,t;0)=$
$$N(t)\:{\bigg[}\delta(p)\:+\:\frac{\mu}{\gamma}\:
M(p;\sigma_{\textrm{m}})\:\frac{1-\exp(-\:\gamma
p^2t/\tau}{p^2}{\bigg]} \eqno(43{\mbox{a}})$$ with the normalizing
coefficient $$N(t)={\bigg\lbrace}1\:+\:
\frac{\mu}{\gamma\sigma_{\textrm{m}}^2}\:{\bigg[}{\bigg(}1\:+\:
\frac{2\gamma\sigma_{\textrm
{m}}^2t}{\tau}{\bigg)}^{1/2}\:-\:1{\bigg]}
{\bigg\rbrace}^{-1}\:.\eqno(44{\mbox{a}})$$ From the probabilistic
viewpoint, the function $N(t)$ is well understandable. When
counting all bacteria living at the time instant $t$, we have to
distinguish whether they are carriers of the original phenotypic
value $p_0=0$ or whether they carry other values, $p\:\neq\:0$.
Since $\int_{-\epsilon}^\epsilon{\mathit\Phi}(p,t;0)=N(t)$ (if
$\epsilon\:\to\:+0$), we may say that a randomly chosen bacterium
may be the carrier of the value $p_0=0$ with the probability equal
to $N(t)$. If $t\:\to\:\infty$, the probability $N(t)$ decreases
towards zero. However, this decreasing $-$ the process influenced
both by the mutations and by the fitness of the bacteria to live
in their environment $-$ is slow. Indeed, let us take
$\gamma=0.02$, $\sigma_{\textrm{m}}=0.05$ and $\mu=10^{-5}$. Then
$\mu/(\gamma\sigma_{\textrm{m}}^2)=0.2$ and
$2\gamma\sigma_{\textrm{m}}^2t/\tau=1$ for the generation number
$t/\tau=10^4$. If these values of $\gamma$ and $\sigma_{\textrm
{m}}$, together with the value 20 minutes for the generation time
$\tau$, may be taken as realistic for some soil bacteria, the
total time $t$ comprising the lifetime of ten thousand generations
of these bacteria equals about five months. If the time $t$ is
roughly ten times (or more than ten times) shorter, formula (44a)
can be simplified:
\begin{equation}\label{shorttime}N(t)\:\approx\:{\bigg(}1\:+\: \frac{\mu
t}{\tau} {\bigg)}^{-1}\ \textrm{ if }\
\frac{2\gamma\sigma_{\textrm
{m}}^2t}{\tau}\:\ll\:1\:.\end{equation} As a rule, the mutation
probability $\mu$ is smaller than $2\gamma\sigma_{\textrm{m}}^2$.
Thus, we may write $$N(t)\:\approx\:1\:-\:\frac{\mu t}{\tau}\
\textrm{ if }\
\frac{2\gamma\sigma_{\textrm{m}}^2t}{\tau}\:\ll\:1\:.\eqno(45')$$

\subsection{Distribution function ${\mathit\Phi}(p,t;0)$ in the
model where $c(p)=\gamma[1-\exp(-\:ap^2)]$}

\bigskip
Now we consider a small parameter $\gamma$ ($0<\gamma\:\ll\:1$)
and another parameter, $a>0$, which need not be small. Only if
$a\:\ll\:1/\sigma_{\textrm{m}}^2$, we may accept the approximation
$c(p)\:\approx\:\gamma ap^2$ and there is no essential difference
from the Waxman-Peck model, only $\gamma$ is replaced by $\gamma
a$.

Otherwise, if $a\sigma_{\textrm{m}}^2$ is comparable with unity,
the integration of the function $\varphi(p,t)$ with respect to $p$
is much more complicated but can be accomplished explicitly. (It
is presented in Appendix.)

Here we confine ourselves to discussing what comes about if
$\:a\sigma_{\textrm{m}}^2\:\gg\:1$. Essentially, under this
condition, we may approximate $1-\exp(-\:ap^2)$ by unity. Then we
obtain the simple result $$\int_{-\infty}^\infty{\mbox
{d}}p\:\varphi(p,t)\:\approx\:1\:+\:
\frac{\mu}{\gamma}\:{\bigg[}1\:-\: \exp{\bigg(}-\:\frac{\gamma
t}{\tau}{\bigg)}{\bigg]}\:.$$ Correspondingly, if
$a\sigma_{\textrm{m}}^2\:\gg\:1$, then

\bigskip
${\mathit\Phi}(p,t;0)\:\approx\:$
$$N(t)\:{\bigg\lbrace}\delta(z)\:+\:
\frac{\mu}{\gamma}\:M(p;\sigma_{\textrm{m}})\:{\bigg[}1-
\exp{\bigg(}-\:\frac{\gamma
t}{\tau}{\bigg)}{\bigg]}{\bigg\rbrace}\:, \eqno(43{\mbox{b}})$$
where
$$N(t)={\bigg\lbrace}1\:+\:\frac{\mu}{\gamma}\:{\bigg[}1\:-\:
\exp{\bigg(}-\:\frac{\gamma
t}{\tau}{\bigg)}{\bigg]}{\bigg\rbrace}^{-1}\:.
\eqno(44{\mbox{b}})$$ In the short-time approximation, formulae
(\ref{shorttime}) and ($45'$) are equally valid as in the case A.
Note that expression (43b) for the distribution function
${\mathit\Phi}(p,t)$ would be correct if $c(p)=1-w(p)/w(0)$ might
be approximated by a small constant $\gamma>0$. In this case,
$N(t)$ may again be approximated as $1-\mu t/\gamma$ at short
enough times. However, if $t\:\to\:\infty$, then $N(t)$ does not
tend to zero (in contrast to the case analyzed in the preceding
subsection):
$$\lim_{t\:\to\:\infty}\:N(t)=\frac{\gamma}{\gamma+\mu}\:.$$

\section{Concluding remarks}

\bigskip
In the present paper, we have focused attention on the importance
of the fitness function $w(p)$ in the theory of the population
genetics. Assuming that $0<c(p)=1-w(p)/w(0)\:\ll\:1$, we have
essentially followed  Waxman and Peck who derived the distribution
function ${\mathit\Phi}(p,t)$ of the population genetics as a
functional of a function $\varphi(p,t)$ (cf. expression
(\ref{Waxsubst})) satisfying a {\textit{linear}}
integro-differential equation (cf. equation (\ref{Waxdiflin})).
However, in contrast with paper [12] where $c(p)$ was approximated
as $\gamma p^2$ with some small parameter $\gamma>0$, we emphasize
that $c(p)$ may be chosen from a wider class of functions. In
particular, we have dealt with the model defined by the function
$c(p)=\gamma[1-\exp(-ap^2)]$.

We have calculated the distribution function as a series with
respect to the mutation probability $\mu$. Our iteration scheme
for calculating the Green function $G(p,t;p_0)$ of the equation
for $\varphi(p,t)$ has been used in the same manner as in the
density-matrix theory.

The replacement of $c(p)$ by $E(p)$, $t$ by $\beta$ (with
$\hbar=1$) and $G_0(p,t;p_0)$ by the unperturbed canonical density
matrix $C_\beta^{(0)}(p,p_0)$ yields the equation
\begin{equation}\label{Bloch0}-\:
\frac{\partial C_\beta^{(0)}(p,p_0)}{\partial\beta}=
E(p)\:C_\beta^{(0)}(p,p_0)\:.\end{equation} With adequately chosen
function $E(p)$, this equation may concern conduction electrons in
a homogeneous non-degenerate semiconductor. (Since $E(p)$ is not
equal to the kinetic energy of an electron in vacuum, we may
interpret the conduction electrons as quasiparticles defined by
the dispersion law $E=E(p)$.)

Our second remark concerns the analogy with the diffusion theory.
The Fourier transform of the function $G(p,t;p_0)$ (multiplied by
a constant) can be interpreted as the concentration $C(x,t;x_0)$
of diffusants which all were initially, at the time $t_0=0$,
localized in the point $x_0$. In the approximation of the present
paper, we may generally write the equation $$\frac{\partial
C_0(x,t;x_0)}{\partial t}=\frac{1}{\tau}\:c\bigg(
-\:{\mbox{i}}\frac{\partial}{\partial x}\bigg) \:C_0(x,t;x_0)\:$$
\begin{equation}\label{Ceqgen}+\:\frac{\mu}{\tau}\:
\exp{\bigg(}-\:\frac{\sigma_{\textrm
{m}}^2x^2}{2}{\bigg)}C(x,t;x_0)\:.\end{equation} If $c(p)=\gamma
p^2$, the concentration $C(x,t;x_0)$ obeys the usual diffusion
equation with the diffusion coefficient $D=\gamma/\tau$. If
$c(p)\:\neq\:\gamma p^2$, the diffusion is anomalous. In any case,
the positiveness of the `potential-energy term' means that
equation (\ref{Ceqgen}) involves a {\textit{creation}} of
diffusants.

If $\mu=0$, we observe that
$N_0=\int_{-\infty}^\infty{\mbox{d}}x\:C_0(x,t;x_0)$ is a quantity
not varying in time. Therefore, we may define the probability
density $P_0(x,t;x_0)=C_0(x,t;x_0)/N_0$ and put the theory on an
equal footing with the theory of the Brownian motion.

If $c(p)=\gamma p^2$ and $\mu=0$, we may write down the Langevin
equation ${\dot x}(u)=(2\gamma/\tau)^{1/2}{\tilde f}(u)$ for the
stochastic paths $x(u)$ ($0\:\leq\:u\:\leq\:t$) which all start
from the common point $x(0)=x_0$ at the time instant $u_0=0$. The
value of the end-point $x(t)=x$ at a given time instant $t>0$ may
be arbitrary and $P_0(x,t;x_0)={\langle}\delta(x-x(t)){\rangle}$.
In the terminology of the theory of stochastic processes, $x(u)$
is the Wiener process.

But then a natural question arises: which stochastic process does
correspond to the case when the fitness function $w(p)$ is modeled
by function (8b) with which we have exemplified our problem? About
ten years ago, we dealt with the equation $$\frac{\partial
P_0(x,t;x_0)}{\partial t}=\frac{\gamma}{\tau}\:
{\bigg[}\exp{\bigg(}a\frac{\partial^2}{\partial
x^2}{\bigg)-1}{\bigg]}\: P_0(x,t;x_0)\:$$
\begin{equation}{\label{Cmultip}}+\:D_0\:
\frac{\partial^2P_0(x,t;x_0)}{{\partial}x^2}\:.\end{equation} (Cf.
equation (45) in [18]; see also [19].) Equation (\ref{Cmultip})
corresponds to a stochastic process with paths $x(u)$ defined by
the stochastic equation ${\dot
x}(u)=[(2D_0)^{1/2}+a\sum_{j}\delta(u-u_j)]\:{\tilde f}(u)$, where
${\tilde f}(u)$ is the standard zero-centered Gaussian white-noise
function and where the sum represents a point process in which
$u_j$ are random time instants distributed in the Poissonian way.
The Poissonian process consists of equal delta-pulses: all the
pulses are taken with the same amplitude $a$. The average
frequency of these pulses is equal to $\gamma/\tau$. Clearly, we
consider a {\textit{multiplicative}} stochastic process $x(u)$
($0\:\leq\:u\:\leq\:t$). If $D_0=0$, then the probability density
$P_0(x,t;x_0)={\langle}\delta(x-x(t)){\rangle}$ is the fundamental
solution of equation (\ref{Cmultip}). Alternatively (as we have
shown in [18]), equation (\ref{Cmultip}) can be written in the
equivalent integro-differential form: $$\frac{\partial
P_0(x,t;x_0)}{\partial t}=\frac{\gamma}{\tau}\:
\int_{-\infty}^\infty{\mbox {d}}x'{\bigg[}\frac{1}{(2\pi
a)^{1/2}}\: \exp{\bigg(}-\:\frac{(x-x')^2}{2a}{\bigg)}$$
\begin{equation}{\label{Cintdif}}-\ \delta(x-x'){\bigg]}
P_0(x,t;x_0)\:+\:D_0\:\frac{\partial^2P_0(x,t;x_0)}{\partial
x^2}\:.\end{equation} In the case when $D_0=0$, equation
(\ref{Cintdif}) was employed by Laskin [20] in a theory of the
channeling of high-energy particles in crystals. (The channeling
occurs when a ray of equi-energy particles bombarding a crystal is
collimated very precisely in a favorable direction.)

In the framework of the diffusion theory, we may conclude that the
parameter $\gamma$ of the theory of the population genetics
corresponds to an environmental noise. If $\gamma=0$, the noise is
absent.

Section IV of the present paper has been devoted to the problem of
the evolution of a population in which all individuals are
initially equal, being the carriers of the phenotypic value
$p_0=0$. The distribution function ${\mathit\Phi}(p,t;0)$ of the
population is a sum of a sharp delta-function component,
$N(t)\delta(p)$, and a blurred component. Similarly as in the
thermodynamics, we may distinguish two phases in the population at
any time $t>0$. Let us denote them as phase S and phase B. The
phase S consist of the carriers of the initial phenotypic value
$p_0=0$. The phase B consists of the individuals carrying the
phenotypic values $p\:\neq\:p_0$. In the Waxman-Peck model (cf.
expressions (43a) and (44a)), the probability $N(t)$ tends to zero
if $t\:\to\:\infty$. Therefore, we may say that the phase S
dissolves gradually in the phase B. In the model with the fitness
function $w(p)$ defined by expression (8b) (or by another similar
expression), the probability $N(t)$ does not tend asymptotically
to zero: this model predicts that both the phases S and B may
coexist if $t\:\to\:\infty$.

\begin{acknowledgments}
This work has been supported by the Grant Agency VEGA of the
Slovak Academy of Sciences and of the Ministry of Education of the
Slovak Republic under contract No. 1/7656/20. I thank D. Waxman
for sending me some reprints of his papers. I thank R. Hlubina and
A. Plecenik for their critical reading of my manuscript. I thank
also two referees for valuable notes to the first version of the
present paper.
\end{acknowledgments}

\appendix*\section{}
In the model where $c(p)=\gamma\:[1-\exp(-\:ap^2)]$, we have to
manage the function $$\frac{1-\exp[-\:c(p)t/\tau]}{c(p)}=$$
$$\frac{1-\exp{\lbrace}-\:\gamma[1-\exp(-\:ap^2)]t/\tau{\rbrace}}
{\gamma[1-\exp(-\:ap^2)]}$$ $$=\frac{1}{\tau}\:\int_0^t{\mbox
{d}}t_1\:
\exp{\bigg(}-\:\frac{\gamma[1-\exp(-\:ap^2)]\:t_1}{\tau}{\bigg)}$$
$$=\frac{1}{\tau}\:\int_0^t{\mbox{d}}t_1\:
\exp{\bigg(}-\:\frac{\gamma t_1}{\tau}{\bigg)}\:
\exp{\bigg(}\frac{\gamma t_1}{\tau}\:\exp(-\:ap^2){\bigg)}\:.$$ We
have to calculate the integral $$\int_{-\infty}^\infty{\mbox
{d}}p\:\varphi(p,t)=1\:+\:\frac{\mu}{\tau}\: \int_0^t{\mbox
{d}}t_1\:\exp{\bigg(}-\:\frac{\gamma t_1}{\tau}{\bigg)}\:I(t_1)
\:,$$ where $$I(t_1)={\bigg(}\frac{1}{2\pi\sigma_{\textrm
{m}}^2}{\bigg)}^{1/2}\int_{-\infty}^\infty{\mbox{d}}p\:
\exp{\bigg(}-\:\frac{p^2}{2\sigma_{\textrm{m}}^2}{\bigg)}\:$$
$$\times\ \exp{\bigg(}\frac{\gamma
t_1}{\tau}\:\exp(-\:ap^2){\bigg)}\:.$$ When developing the second
exponential in the MacLaurin series, we obtain the following sum
of the Laplace integrals:

\bigskip
$I(t_1)=$ $${\bigg(}\frac{1}{2\pi\sigma_{\textrm
{m}}^2}{\bigg)}^{1/2}\: \sum_{j=0}^\infty\:
\frac{t_1^j}{j!\:\tau^j}\:\int_{-\:\infty}^\infty{\mbox{d}}p\:
\exp{\bigg[}-\:{\bigg(}\frac{1}{2\sigma_{\textrm
{m}}^2}\:+\:ja{\bigg)}\:p^2 {\bigg]}\:.$$ Hence,
$$I(t_1)=\sum_{j=0}^\infty\:
\frac{t_1^j}{j!\:\tau^j\:(1+2ja\sigma_{\textrm{m}}^2)^{1/2}}\:.$$
In this way we have obtained the result

\bigskip
$\int_{-\:\infty}^\infty{\mbox {d}}p\:\varphi(p,t)=$
$$1\:+\:\frac{\mu}{\tau}\: \sum_{j=0}^\infty\:
\frac{1}{j!\:\tau^j\:(1+2ja\sigma_{\textrm{m}}^2)^{1/2}}
\int_0^t{\mbox{d}}t_1\:\exp{\bigg(}-\:\frac{\gamma
t_1}{\tau}{\bigg)}\:t_1^j \:.$$ The integral in the r.h. side of
this equality is easily calculable: $$\int_0^t{\mbox
{d}}t_1\:\exp{\bigg(}-\:\frac{\gamma t_1}{\tau}{\bigg)}\:t_1^j=$$
$$\frac{j!\:\tau^{j+1}}{\gamma^{j+1}}\:{\bigg[}1\:-\:
\exp{\bigg(}-\:\frac{\gamma t}{\tau}{\bigg)}\:\sum_{k=0}^j\:
\frac{\gamma^kt^k}{k!\:\tau^k}{\bigg]}\:.$$ Thus we have obtained
the distribution function
$${\mathit\Phi}(p,t;0)=N(t)\:{\bigg[}\delta(p)\:$$
$$+\:\frac{\mu}{\gamma}\: M(p;\sigma_{\textrm{m}})\:
\frac{1-\:\exp{\lbrace}-\:\gamma[1-\exp(-\:ap^2]t/\tau}
{1-\exp(-\:ap^2t/\tau)}{\bigg]}\:,$$ where

$$N(t)={\bigg\lbrace}1\:+\:\frac{\mu}{\gamma}\ \times$$
$$\sum_{j=0}^\infty\: \frac{1}{\gamma^j(1+2ja\sigma_{\textrm
{m}}^2)^{1/2}}\:{\bigg[}1\:-\: \exp{\bigg(}-\:\frac{\gamma
t}{\tau}{\bigg)}\sum_{k=0}^j\:
\frac{\gamma^kt^k}{k!\:\tau^k}{\bigg]}{\bigg\rbrace}^{-1}\:.$$

\section*{References}

{\flushleft{[1] Chandrasekhar S., Revs. Mod. Phys. {\textbf{15}}
(1943) 2 (also in: {\textit{Selected Papers on Noise and
Stochastic Processes}}, Ed.: Wax N., Dover, New York 1954, 3)}}
\newline
[2] Kampen N. G. van, {\textit{Stochastic Processes in Physics and
Chemistry}}, North Holland, Amsterdam 1981
\newline
[3] Feynman R. P., {\textit{Statistical Mechanics}}, W. A.
Benjamin, Reading 1972
\newline
[4] Feynman R. P., Hibbs A. R., {\textit{Quantum Mechanics and
Path Integrals}}, McGraw Hill, New York 1965
\newline
[5] Bez{\'a}k V., Acta Physica Slovaca {\textbf{28}} (1978), 12;
Acta Physica Slovaca {\textbf{28}} (1978) 24
\newline
[6] Davis H., {\textit{Introduction to Nonlinear Differential and
Integral Equations}}, Dover, New York 1962
\newline
[7] Salem L. D., Montemayor R., Phys. Rev. A {\textbf{43}} (1991)
1162
\newline
[8] Montemayor R., Salem L. D., Phys. Rev. A {\textbf{44}} (1991)
7037
\newline
[9] Salem L. D., Montemayor R., Phys. Rev. A {\textbf{47}} (1993)
105
\newline
[10] Bessis N., Bessis G., J. Math. Phys. {\textbf{38}} (1997)
5483
\newline
[11] Nowakowski M., Rosu H. C., Phys. Rev. E {\textbf{65}} (2002)
047602
\newline
[12] Waxman D., Contemporary Physics {\textbf{43}}
(2002) 13
\newline
[13] Waxman D., Peck J. R., Science {\textbf{279}} (1998) 1210
\newline
[14] Coppersmith S. N., Blank R. D., Kadanoff L. P., J.
Statistical Phys. {\textbf{97}} (1999) 429
\newline
[15] Callaway J., {\textit{Quantum Theory of the Solid State, Part
B}}, Academic Press, New York 1974
\newline
[16] Bez{\'a}k V., J. Math. Phys. {\textbf{37}} (1996) 5939
\newline
[17] March N. H., Young W. H., Sampanthar S., {\textit{The
Many-Body Problem in Quantum Mechanics}}, University Press,
Cambridge 1967
\newline
[18] Bez{\' a}k V., J. Phys. A: Math. Gen. {\textbf{25}} (1992)
6027
\newline
[19] Bez{\' a}k V., Physica A {\textbf{206}} (1994) 127
\newline
[20] Laskin N. V., J. Phys. A: Math. Gen. {\textbf{22}} (1989)
1565
\end{document}